\documentclass[sigconf]{acmart}

\AtBeginDocument{%
  \providecommand\BibTeX{{%
    \normalfont B\kern-0.5em{\scshape i\kern-0.25em b}\kern-0.8em\TeX}}}

\copyrightyear{2024} 
\acmYear{2024} 
\setcopyright{rightsretained} 
\acmConference[MOBILEHCI Adjunct '24]{26th International Conference on Mobile Human-Computer Interaction}{September 30-October 3, 2024}{Melbourne, VIC, Australia}
\acmBooktitle{26th International Conference on Mobile Human-Computer Interaction (MOBILEHCI Adjunct '24), September 30-October 3, 2024, Melbourne, VIC, Australia}
\acmDOI{10.1145/3640471.3686646}
\acmISBN{979-8-4007-0506-9/24/09}

\usepackage{multirow}
\usepackage{graphicx}
\usepackage{subfigure} 
\usepackage{xcolor,soul}
\usepackage{array}
\usepackage{booktabs}
\usepackage{tabularx}

\begin{document}


\title[Towards Enhanced Context Awareness with Vision-based Multimodal Interfaces]{Towards Enhanced Context Awareness with Vision-based Multimodal Interfaces}

\author{Yongquan Hu}
\orcid{0000-0003-1315-8969}
 \email{yongquan.hu@unsw.edu.au}
\affiliation{%
 \institution{PhD Student, School of Computer Science and Engineering, UNSW}
 \country{Australia}}

 \author{Wen Hu}
\orcid{0000-0002-4076-1811}
\email{wen.hu@unsw.edu.au}
\affiliation{%
 \institution{PhD Supervisor, School of Computer Science and Engineering, UNSW}
 \country{Australia}}

 \author{Aaron Quigley}
\orcid{0000-0002-5274-6889}
\email{aquigley@acm.org}
\affiliation{%
 \institution{PhD Supervisor, Data61, CSIRO}
 \country{Australia}}

\renewcommand{\shortauthors}{Yongquan Hu et al.}

\begin{abstract}


Vision-based Interfaces (VIs) are pivotal in advancing Human-Computer Interaction (HCI), particularly in enhancing context awareness. However, there are significant opportunities for these interfaces due to rapid advancements in multimodal Artificial Intelligence (AI), which promise a future of tight coupling between humans and intelligent systems. AI-driven VIs, when integrated with other modalities, offer a robust solution for effectively capturing and interpreting user intentions and complex environmental information, thereby facilitating seamless and efficient interactions. This PhD study explores three application cases of multimodal interfaces to augment context awareness, respectively focusing on three dimensions of visual modality: scale, depth, and time: a fine-grained analysis of physical surfaces via microscopic image, precise projection of the real world using depth data, and rendering haptic feedback from video background in virtual environments.

\end{abstract}



\maketitle

\section{Research Topic}

In the domain of ubiquitous and mobile computing, visual data are essential for converting complex information into explainable and actionable results comparing to compared to other modalities such as radar \cite{poslad2011ubiquitous,quigley2000fade,elvitigala2023radarfoot}. The prevalence of Vision-based Interfaces (VIs) in these systems underscores its critical role in enabling intelligent system interactions with both physical and digital environments, which is key for enhancing context awareness \cite{haque2020illuminating,makkonen2009context}. The effectiveness of an interface relies on its ability to perceive, understand, and dynamically respond to its ambient context, making VIs vital for devices to adapt smartly and interact seamlessly with their surroundings \cite{kolsch2004vision,hasan2012human}.

Building on this foundation, Vision-based Multimodal Interfaces (VMIs) could further enhance context awareness by combining with diverse sensory data streams—not just visual cues—into a cohesive understanding of the environment \cite{shin2007vision,gellersen2002multi}. This richer understanding of the environment allows systems to provide more accurate, timely, and relevant responses to user inputs and environmental changes, overcoming the limitations of single-modality systems \cite{sharma1998toward,ismail2015vision}. However, it remains widely unexplored for VMIs, particularly with the advent of integrating them with the emerging multimodal AI technologies, such as Large-scale Language Models (LLMs) capable of processing text, audio, image, and code modalities \cite{tong2024cambrian}. 

To address these challenges, I aim to develop customized VMIs as input/output techniques for diverse scenarios to enhance context awareness, thereby improving the efficiency, friendliness, and accessibility of interactions between devices and users. My PhD study represents an incremental advancement built upon prior research. Specifically, I propose multimodal-AI-driven VMIs that focus on three crucial visual dimensions — \textbf{Scale}, \textbf{Space}, and \textbf{Time} — each essential for developing comprehensive multimodal systems:

\begin{itemize}
\item $1^{st}$ Application - microscopic \textit{Scale} for surface sensing (input): Surface sensing, crucial for context awareness of user behavior and device location, can be enhanced by using microscopic-scale color RGB images for more intuitive and detailed insights.
\item $2^{nd}$ Application - \textit{Spatial} sensitivity of projection creation (input \& output): Accurate depth information enhances spatial interactions, especially in augmented reality settings, where real-world spatial congruence is necessary for engaging with digital content.
\item $3^{rd}$ Application - \textit{Temporal} dynamics in haptic feedback (output): Effective management of environmental effects' temporal dynamics is crucial for real-time responsive applications, such as interactive gaming or dynamic simulations, where delays can impair the user experience.
\end{itemize}

\section{Research Progress} \label{progress}

As summarized in Table \ref{tab: 3 applications}, the three central dimensions underscore my tailored approach to VMIs, spanning applications from surface sensing to tactile feedback rendering. This strategic focus enables the design of adaptive interfaces that enhance context awareness tailored to the specific needs of users in various scenarios. Additionally, this enhanced context awareness serves as a conduit between digital and physical worlds, fostering more intuitive and seamless interactions. My three key projects illustrate this bridging function across a broad spectrum of applications that straddle virtual and physical realms.

So far, The core component of the first application, MicroCam \cite{hu2023microcam}, has been published as a full-length paper at the Proceedings of the ACM on Interactive, Mobile, Wearable and Ubiquitous Technologies (IMWUT) 2023, illustrating the application of the scale dimension. Complementary work related to MicroCam has been submitted to the ACM International Conference on Mobile Human-Computer Interaction (MobileHCI) 2024 Late-Breaking Work (LBW) track. The projects focusing on the second (spatial dimension) and third (temporal dimension) applications have been submitted as full papers to IMWUT 2024 and the ACM Symposium on Virtual Reality Software and Technology (VRST) 2024, respectively. Additionally, related workshops and poster presentations have been published \cite{hu2023exploring, hu2023genair,hu2023towards}. Therefore, I will use the first application as an example to comprehensively address the aforementioned questions. In particular, the research questions for the first application are: (1) \textit{How can microscopic-scale color RGB images be utilized for surface sensing, specifically for object and material recognition?} (2) \textit{Besides visual modality, what other modalities can be employed to surface sensing for enhancing context awareness?}

\begin{table*}[h]
\centering
\caption{Overview of the three key applications in my PhD research}
\resizebox{\linewidth}{!}{
\begin{tabular}{l|l|l|l|l}
\toprule
\textbf{ No.} & \textbf{Application} & \textbf{Context Information Included} & \textbf{Input Modalities} & \textbf{Environment Attributes} \\
\hline
1 & surface sensing & surface material, method description & visual, IMU  & physical \\

2 & projection content creation & spatial parameter, user prompt & visual, speech (textual) & physical + virtual \\

3 & haptic feedback  rendering & particle motion, explosion & visual, auditory & virtual \\
\bottomrule
\end{tabular}
}
\label{tab: 3 applications}
\end{table*}

\section{ Methodology}\label{methodology}

Despite the diverse scenarios and research focuses across my three projects, I consistently address several key aspects that serve as a general template for designing, implementing, and evaluating my work. These aspects include: (1) leveraging the relevant visual dimension—scale, space, or time—in each scenario to maximize context awareness; (2) enhancing context awareness by integrating the visual modality with other modalities; (3) rigorously evaluating these enhancements through systematic testing and feedback analysis. This structured approach ensures a comprehensive understanding of each application's impact and effectiveness.

\begin{figure*}
\centering
  \includegraphics[width=0.75\textwidth]{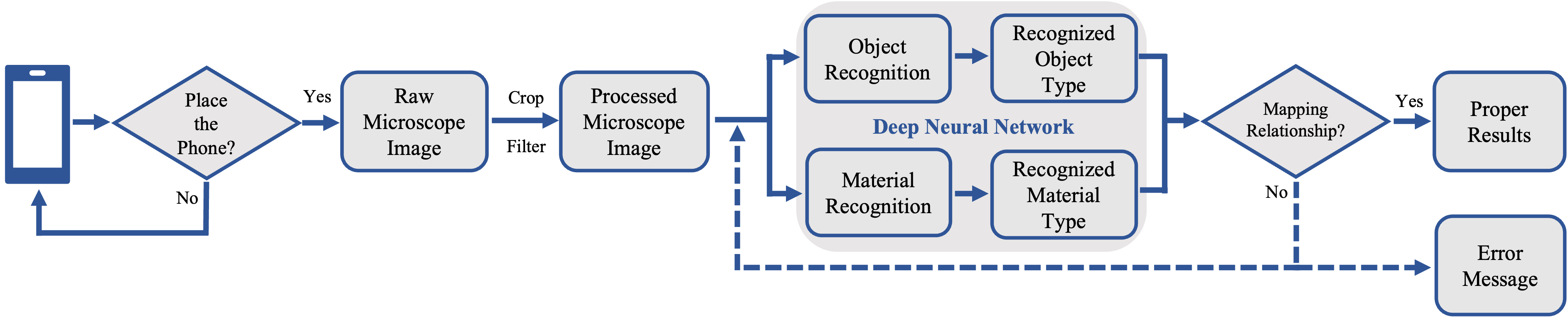}
  \caption{ The schematic representation of the MicroCam system pipeline.}
  ~\label{fig:MicroCam}
\end{figure*}

Specifically, in my first application, MicroCam employs a smartphone's microscope camera to capture fine details of surfaces, enhancing interactive possibilities by focusing on the microscopic scale of visual modality crucial for discerning rich texture and material properties of everyday objects. As illustrated in Figure \ref{fig:MicroCam}, the system uses the smartphone's Inertial Measurement Unit (IMU) to ensure stability before capturing high-resolution images, which are processed using an optimized deep neural network based on MobileNet to identify material characteristics, thereby facilitating smarter interactions. This system incorporates continual learning (experience replay) to enhance optimization and improve the algorithm's generalization and robustness.

Building on MicroCam, the MultiSurf-GPT framework utilizes multimodal data inputs, including radar \cite{yeo2018exploring}, microscope \cite{hu2023microcam}, and multispectral \cite{yeo2017specam} data, processed through GPT-4o's capabilities. Transitioning from single to multimodal inputs, MultiSurf-GPT adopts a unified approach to surface sensing, applying prompt engineering strategies (zero-shot and one-shot prompting) to extract and interpret high-level, context-aware insights from various sensory modalities. This approach deepens environmental understanding and fosters the development of more adaptive and intelligent context-aware applications.

\section{Evaluation}
I will use the first application as an example to illustrate which evaluation methods are applied to evaluate the performance of a technology or system.

\subsection{Quantitative Evaluation}
MicroCam and MultiSurf-GPT employ several quantitative methods to evaluate the efficacy of their surface sensing systems. MicroCam utilizes two primary evaluation approaches: "Time-Split" and "Leave-One-Person-Out" methods, focusing on object and material recognition rates, which highlight the system's robustness and generalization capabilities on unencountered surfaces . These methods effectively test the system's ability to perform under realistic scenarios where the testing dataset has not been seen during training, thus assessing true performance beyond controlled laboratory conditions. In contrast, MultiSurf-GPT employs accuracy metrics across tasks like Rotation, Orientation, and Identification with impressive rates achieved through its use of the Random Forest model and GPT-4o's capabilities in handling complex datasets, such as CSV files from the Tangible Radar \cite{yeo2018exploring} dataset, and image recognition tasks in the MicroCam and SpeCam \cite{yeo2017specam} datasets . Both systems demonstrate a strong reliance on data-driven evaluation metrics to ensure their models' effectiveness across varied real-world applications.

\subsection{Qualitative evaluation}
In the qualitative evaluation, MicroCam's methodology includes user experience analyses that compare its technology against similar technologies like MagicFinger \cite{yang2012magic} and SpectroPhone \cite{schrapel2021spectrophone}, focusing on practical usability aspects such as ease of integration into everyday devices and the intuitive nature of the interaction with the technology . This approach helps in understanding the relative advantages or limitations from a user-centric perspective. Similarly, MultiSurf-GPT's qualitative evaluation delves into context-aware reasoning capabilities, utilizing LLMs to process and interpret multimodal surface sensing data. This is particularly noted in its ability to synthesize information across different sensory inputs, providing enriched contextual insights that could potentially transform user-device interaction in practical scenarios . The qualitative assessments for both MicroCam and MultiSurf-GPT aim to elucidate the technologies' real-world applicability and integration, emphasizing user interaction and the technology's adaptability to everyday use.

\section{Expected Contributions}
The main goal of my PhD study is to explore novel applications of VMIs that have not yet been widely investigated. I focus on three such applications and aim to contribute in the following areas: artifact development, comprehensive surveys, and empirical research, as outlined below.

\begin{itemize}
\label{tab: contributions}
\item Artifact Contributions.
    \begin{itemize}
        \item Development of a new framework of context-aware VMIs with focusing on three dimensions: scale, space and time.
        \item A detection technology utilizing a single visual modality to detect surface materials and object types based on microscope images.
        \item A content-creation technique combining visual modality and textual modality, leveraging a depth camera and contrastive learning to strengthen physical environment perception and understanding of input prompts in projection scenario. 
        \item A haptic rendering method integrating visual modality and auditory modality, converting the background elements of VR video like particle motion and explosion sounds into somatosensory-based haptic feedback, thereby advancing user immersion. 
    \end{itemize}

\item Survey Contribution.
    \begin{itemize}
    \item Comprehensive literature review, providing a taxonomy of context-aware VMIs.
    advancing user immersion. 
    \end{itemize}

\item Empirical Contributions.
    \begin{itemize}
        \item Empirical studies to evaluate the techniques and interactions of developed applications.
        \item Design considerations for vision-based multimodal interfaces, providing new insights and practical solutions to enhance context awareness.
    \end{itemize}

\end{itemize}

\bibliographystyle{ACM-Reference-Format}
\bibliography{sample}


\begin{thebibliography}{20}


\ifx \showCODEN    \undefined \def \showCODEN     #1{\unskip}     \fi
\ifx \showDOI      \undefined \def \showDOI       #1{#1}\fi
\ifx \showISBNx    \undefined \def \showISBNx     #1{\unskip}     \fi
\ifx \showISBNxiii \undefined \def \showISBNxiii  #1{\unskip}     \fi
\ifx \showISSN     \undefined \def \showISSN      #1{\unskip}     \fi
\ifx \showLCCN     \undefined \def \showLCCN      #1{\unskip}     \fi
\ifx \shownote     \undefined \def \shownote      #1{#1}          \fi
\ifx \showarticletitle \undefined \def \showarticletitle #1{#1}   \fi
\ifx \showURL      \undefined \def \showURL       {\relax}        \fi
\providecommand\bibfield[2]{#2}
\providecommand\bibinfo[2]{#2}
\providecommand\natexlab[1]{#1}
\providecommand\showeprint[2][]{arXiv:#2}

\bibitem[Elvitigala et~al\mbox{.}(2023)]%
        {elvitigala2023radarfoot}
\bibfield{author}{\bibinfo{person}{Don~Samitha Elvitigala}, \bibinfo{person}{Yunfan Wang}, \bibinfo{person}{Yongquan Hu}, {and} \bibinfo{person}{Aaron~J Quigley}.} \bibinfo{year}{2023}\natexlab{}.
\newblock \showarticletitle{RadarFoot: Fine-grain Ground Surface Context Awareness for Smart Shoes}. In \bibinfo{booktitle}{\emph{Proceedings of the 36th Annual ACM Symposium on User Interface Software and Technology}}. \bibinfo{pages}{1--13}.
\newblock


\bibitem[Gellersen et~al\mbox{.}(2002)]%
        {gellersen2002multi}
\bibfield{author}{\bibinfo{person}{Hans~W Gellersen}, \bibinfo{person}{Albrecht Schmidt}, {and} \bibinfo{person}{Michael Beigl}.} \bibinfo{year}{2002}\natexlab{}.
\newblock \showarticletitle{Multi-sensor context-awareness in mobile devices and smart artifacts}.
\newblock \bibinfo{journal}{\emph{Mobile Networks and Applications}}  \bibinfo{volume}{7} (\bibinfo{year}{2002}), \bibinfo{pages}{341--351}.
\newblock


\bibitem[Haque et~al\mbox{.}(2020)]%
        {haque2020illuminating}
\bibfield{author}{\bibinfo{person}{Albert Haque}, \bibinfo{person}{Arnold Milstein}, {and} \bibinfo{person}{Li Fei-Fei}.} \bibinfo{year}{2020}\natexlab{}.
\newblock \showarticletitle{Illuminating the dark spaces of healthcare with ambient intelligence}.
\newblock \bibinfo{journal}{\emph{Nature}} \bibinfo{volume}{585}, \bibinfo{number}{7824} (\bibinfo{year}{2020}), \bibinfo{pages}{193--202}.
\newblock


\bibitem[Hasan and Kareem(2012)]%
        {hasan2012human}
\bibfield{author}{\bibinfo{person}{Haitham~Sabah Hasan} {and} \bibinfo{person}{S~Abdul Kareem}.} \bibinfo{year}{2012}\natexlab{}.
\newblock \showarticletitle{Human computer interaction for vision based hand gesture recognition: A survey}. In \bibinfo{booktitle}{\emph{2012 International Conference on Advanced Computer Science Applications and Technologies (ACSAT)}}. IEEE, \bibinfo{pages}{55--60}.
\newblock


\bibitem[Hu et~al\mbox{.}(2023a)]%
        {hu2023towards}
\bibfield{author}{\bibinfo{person}{Yongquan Hu}, \bibinfo{person}{Wen Hu}, {and} \bibinfo{person}{Aaron Quigley}.} \bibinfo{year}{2023}\natexlab{a}.
\newblock \showarticletitle{Towards Using Generative AI for Facilitating Image Creation in Spatial Augmented Reality}. In \bibinfo{booktitle}{\emph{2023 IEEE International Symposium on Mixed and Augmented Reality Adjunct (ISMAR-Adjunct)}}. IEEE, \bibinfo{pages}{441--443}.
\newblock


\bibitem[Hu et~al\mbox{.}(2023b)]%
        {hu2023microcam}
\bibfield{author}{\bibinfo{person}{Yongquan Hu}, \bibinfo{person}{Hui-Shyong Yeo}, \bibinfo{person}{Mingyue Yuan}, \bibinfo{person}{Haoran Fan}, \bibinfo{person}{Don~Samitha Elvitigala}, \bibinfo{person}{Wen Hu}, {and} \bibinfo{person}{Aaron Quigley}.} \bibinfo{year}{2023}\natexlab{b}.
\newblock \showarticletitle{Microcam: Leveraging smartphone microscope camera for context-aware contact surface sensing}.
\newblock \bibinfo{journal}{\emph{Proceedings of the ACM on Interactive, Mobile, Wearable and Ubiquitous Technologies}} \bibinfo{volume}{7}, \bibinfo{number}{3} (\bibinfo{year}{2023}), \bibinfo{pages}{1--28}.
\newblock


\bibitem[Hu et~al\mbox{.}(2023c)]%
        {hu2023exploring}
\bibfield{author}{\bibinfo{person}{Yongquan Hu}, \bibinfo{person}{Mingyue Yuan}, \bibinfo{person}{Kaiqi Xian}, \bibinfo{person}{Don~Samitha Elvitigala}, {and} \bibinfo{person}{Aaron Quigley}.} \bibinfo{year}{2023}\natexlab{c}.
\newblock \showarticletitle{Exploring the design space of employing ai-generated content for augmented reality display}.
\newblock \bibinfo{journal}{\emph{arXiv preprint arXiv:2303.16593}} (\bibinfo{year}{2023}).
\newblock


\bibitem[Hu et~al\mbox{.}(2023d)]%
        {hu2023genair}
\bibfield{author}{\bibinfo{person}{Yongquan Hu}, \bibinfo{person}{Dawen Zhang}, {and} \bibinfo{person}{Aaron Quigley}.} \bibinfo{year}{2023}\natexlab{d}.
\newblock \showarticletitle{GenAIR: Exploring Design Factor of Employing Generative AI for Augmented Reality}. In \bibinfo{booktitle}{\emph{Proceedings of the 2023 ACM Symposium on Spatial User Interaction}}. \bibinfo{pages}{1--3}.
\newblock


\bibitem[Ismail et~al\mbox{.}(2015)]%
        {ismail2015vision}
\bibfield{author}{\bibinfo{person}{Ajune~Wanis Ismail}, \bibinfo{person}{Mark Billinghurst}, {and} \bibinfo{person}{Mohd~Shahrizal Sunar}.} \bibinfo{year}{2015}\natexlab{}.
\newblock \showarticletitle{Vision-based technique and issues for multimodal interaction in augmented reality}. In \bibinfo{booktitle}{\emph{Proceedings of the 8th International Symposium on Visual Information Communication and Interaction}}. \bibinfo{pages}{75--82}.
\newblock


\bibitem[Kolsch et~al\mbox{.}(2004)]%
        {kolsch2004vision}
\bibfield{author}{\bibinfo{person}{Mathias Kolsch}, \bibinfo{person}{Matthew Turk}, {and} \bibinfo{person}{Tobias Hollerer}.} \bibinfo{year}{2004}\natexlab{}.
\newblock \showarticletitle{Vision-based interfaces for mobility}. In \bibinfo{booktitle}{\emph{The First Annual International Conference on Mobile and Ubiquitous Systems: Networking and Services, 2004. MOBIQUITOUS 2004.}} IEEE, \bibinfo{pages}{86--94}.
\newblock


\bibitem[Makkonen et~al\mbox{.}(2009)]%
        {makkonen2009context}
\bibfield{author}{\bibinfo{person}{Jarmo Makkonen}, \bibinfo{person}{Ivan Avdouevski}, \bibinfo{person}{Riitta Kerminen}, {and} \bibinfo{person}{Ari Visa}.} \bibinfo{year}{2009}\natexlab{}.
\newblock \showarticletitle{Context awareness in human-computer interaction}.
\newblock In \bibinfo{booktitle}{\emph{Human-Computer Interaction}}. \bibinfo{publisher}{IntechOpen}.
\newblock


\bibitem[Poslad(2011)]%
        {poslad2011ubiquitous}
\bibfield{author}{\bibinfo{person}{Stefan Poslad}.} \bibinfo{year}{2011}\natexlab{}.
\newblock \bibinfo{booktitle}{\emph{Ubiquitous computing: smart devices, environments and interactions}}.
\newblock \bibinfo{publisher}{John Wiley \& Sons}.
\newblock


\bibitem[Quigley and Eades(2000)]%
        {quigley2000fade}
\bibfield{author}{\bibinfo{person}{Aaron Quigley} {and} \bibinfo{person}{Peter Eades}.} \bibinfo{year}{2000}\natexlab{}.
\newblock \showarticletitle{Fade: Graph drawing, clustering, and visual abstraction}. In \bibinfo{booktitle}{\emph{International Symposium on Graph Drawing}}. Springer, \bibinfo{pages}{197--210}.
\newblock


\bibitem[Schrapel et~al\mbox{.}(2021)]%
        {schrapel2021spectrophone}
\bibfield{author}{\bibinfo{person}{Maximilian Schrapel}, \bibinfo{person}{Philipp Etgeton}, {and} \bibinfo{person}{Michael Rohs}.} \bibinfo{year}{2021}\natexlab{}.
\newblock \showarticletitle{SpectroPhone: Enabling Material Surface Sensing with Rear Camera and Flashlight LEDs}. In \bibinfo{booktitle}{\emph{Extended Abstracts of the 2021 CHI Conference on Human Factors in Computing Systems}}. \bibinfo{pages}{1--5}.
\newblock


\bibitem[Sharma et~al\mbox{.}(1998)]%
        {sharma1998toward}
\bibfield{author}{\bibinfo{person}{Rajeev Sharma}, \bibinfo{person}{Vladimir~I Pavlovic}, {and} \bibinfo{person}{Thomas~S Huang}.} \bibinfo{year}{1998}\natexlab{}.
\newblock \showarticletitle{Toward multimodal human-computer interface}.
\newblock \bibinfo{journal}{\emph{Proc. IEEE}} \bibinfo{volume}{86}, \bibinfo{number}{5} (\bibinfo{year}{1998}), \bibinfo{pages}{853--869}.
\newblock


\bibitem[Shin and Chun(2007)]%
        {shin2007vision}
\bibfield{author}{\bibinfo{person}{Gihan Shin} {and} \bibinfo{person}{Junchul Chun}.} \bibinfo{year}{2007}\natexlab{}.
\newblock \showarticletitle{Vision-based multimodal human computer interface based on parallel tracking of eye and hand motion}. In \bibinfo{booktitle}{\emph{2007 International Conference on Convergence Information Technology (ICCIT 2007)}}. IEEE, \bibinfo{pages}{2443--2448}.
\newblock


\bibitem[Tong et~al\mbox{.}(2024)]%
        {tong2024cambrian}
\bibfield{author}{\bibinfo{person}{Shengbang Tong}, \bibinfo{person}{Ellis Brown}, \bibinfo{person}{Penghao Wu}, \bibinfo{person}{Sanghyun Woo}, \bibinfo{person}{Manoj Middepogu}, \bibinfo{person}{Sai~Charitha Akula}, \bibinfo{person}{Jihan Yang}, \bibinfo{person}{Shusheng Yang}, \bibinfo{person}{Adithya Iyer}, \bibinfo{person}{Xichen Pan}, {et~al\mbox{.}}} \bibinfo{year}{2024}\natexlab{}.
\newblock \showarticletitle{Cambrian-1: A Fully Open, Vision-Centric Exploration of Multimodal LLMs}.
\newblock \bibinfo{journal}{\emph{arXiv preprint arXiv:2406.16860}} (\bibinfo{year}{2024}).
\newblock


\bibitem[Yang et~al\mbox{.}(2012)]%
        {yang2012magic}
\bibfield{author}{\bibinfo{person}{Xing-Dong Yang}, \bibinfo{person}{Tovi Grossman}, \bibinfo{person}{Daniel Wigdor}, {and} \bibinfo{person}{George Fitzmaurice}.} \bibinfo{year}{2012}\natexlab{}.
\newblock \showarticletitle{Magic finger: always-available input through finger instrumentation}. In \bibinfo{booktitle}{\emph{Proceedings of the 25th annual ACM symposium on User interface software and technology}}. \bibinfo{pages}{147--156}.
\newblock


\bibitem[Yeo et~al\mbox{.}(2017)]%
        {yeo2017specam}
\bibfield{author}{\bibinfo{person}{Hui-Shyong Yeo}, \bibinfo{person}{Juyoung Lee}, \bibinfo{person}{Andrea Bianchi}, \bibinfo{person}{David Harris-Birtill}, {and} \bibinfo{person}{Aaron Quigley}.} \bibinfo{year}{2017}\natexlab{}.
\newblock \showarticletitle{Specam: Sensing surface color and material with the front-facing camera of a mobile device}. In \bibinfo{booktitle}{\emph{Proceedings of the 19th International Conference on Human-Computer Interaction with Mobile Devices and Services}}. \bibinfo{pages}{1--9}.
\newblock


\bibitem[Yeo et~al\mbox{.}(2018)]%
        {yeo2018exploring}
\bibfield{author}{\bibinfo{person}{Hui-Shyong Yeo}, \bibinfo{person}{Ryosuke Minami}, \bibinfo{person}{Kirill Rodriguez}, \bibinfo{person}{George Shaker}, {and} \bibinfo{person}{Aaron Quigley}.} \bibinfo{year}{2018}\natexlab{}.
\newblock \showarticletitle{Exploring tangible interactions with radar sensing}.
\newblock \bibinfo{journal}{\emph{Proceedings of the ACM on Interactive, Mobile, Wearable and Ubiquitous Technologies}} \bibinfo{volume}{2}, \bibinfo{number}{4} (\bibinfo{year}{2018}), \bibinfo{pages}{1--25}.
\newblock


\end{thebibliography}


\end{document}